\documentclass[conference]{IEEEtran}
\IEEEoverridecommandlockouts
\usepackage{cite}
\usepackage{amsmath,amssymb,amsfonts}
\usepackage{algorithmic}
\usepackage{graphicx}
\usepackage{textcomp}
\usepackage{xcolor}
\def\BibTeX{{\rm B\kern-.05em{\sc i\kern-.025em b}\kern-.08em
    T\kern-.1667em\lower.7ex\hbox{E}\kern-.125emX}}

\usepackage{url}
\usepackage{listings}
\usepackage{verbatim}

\begin{document}

\title{An Open-Source Framework to Emulate Delay and Disruption Tolerant Networks for International Space Station Communication
%\thanks{NSERC USRA}
}

\author{\IEEEauthorblockN{1\textsuperscript{st} Krit Grover}
\IEEEauthorblockA{\textit{Department of Computer and Mathematical Sciences} \\
\textit{University of Toronto Scarborough}\\
Toronto, Canada \\
krit.grover@mail.utoronto.ca}
\and
\IEEEauthorblockN{2\textsuperscript{nd} Marcelo Ponce}
\IEEEauthorblockA{\textit{Department of Computer and Mathematical Sciences} \\
\textit{University of Toronto Scarborough}\\
Toronto, Canada \\
m.ponce@utoronto.ca}
}

\maketitle

\begin{abstract}
Delay and Disruption Tolerant Networks (DTN) are critical for reliable
communications in challenged network environments, particularly for space
systems where end-to-end connectivity cannot be guaranteed. We present an
open-source, full-stack implementation of the Bundle protocol for communicating with the International
Space Station (ISS), with complete security features including Bundle Authentication
Block (BAB), Payload Integrity Block (PIB), and Payload Confidentiality Block (PCB)
using HMAC-SHA256 and AES-256-CBC encryption. The system includes bundle fragmentation
and reassembly, priority-based queuing, custody transfer with ACK/NAK mechanisms,
and automatic retransmission.
Our system also includes a frontend facilitated by a modern responsive web interface. We consider this work highly relevant in the context of
computer networking because:
i) it demonstrates a full stack, open-source, freely available
implementation of this critical and reliable protocol; and
ii) it offers an interactive educational and learning framework in the
field of computer networks and communications.

\end{abstract}


%%%%%%%%%%%%%%%%%%%%%%%%%%%%%%%%%%%%%%%%%%%%%%%%%%%%%%%%%%%
\section{Introduction \& Motivation}
\label{sec:intro}
%%%%%%%%%%%%%%%%%%%%%%%%%%%%%%%%%%%%%%%%%%%%%%%%%%%%%%%%%%%

The Internet's foundational protocols (TCP/IP) were designed under assumptions of continuous end-to-end connectivity, low latency, and symmetric bidirectional links. While these assumptions hold for most terrestrial networks, they fundamentally break down in challenged environments such as deep space communications, disaster recovery scenarios, remote sensor networks, and low-Earth orbit (LEO) satellite systems. The International Space Station (ISS), orbiting at approximately 420 km altitude and traveling at 7.66 km/s, presents a compelling case study: ground station contact windows last only 5--10 minutes per pass, with gaps of 45--90 minutes between successive contacts. Traditional transport protocols like TCP would consider these disconnections as network failures, triggering congestion control mechanisms and ultimately failing to deliver data reliably.

Delay and Disruption Tolerant Networking (DTN) addresses these challenges through a fundamentally different paradigm. Rather than assuming persistent connectivity, DTN uses a ``store-and-forward'' architecture where intermediate nodes persistently store messages (called \textit{bundles}) until a suitable forwarding opportunity arises. The DTN architecture has been endorsed by the Consultative Committee for Space Data Systems (CCSDS) and is actively deployed on the ISS, making it an increasingly important topic in computer networking education.

Despite DTN's growing significance, accessible educational tools that demonstrate these protocols in action and provide a practical implementation remain scarce. Students learning about delay-tolerant networking often encounter only theoretical descriptions without the opportunity to observe bundle routing, custody transfer, link budget dynamics, and handoff procedures in a realistic operational context. As of today, just a handful of examples are available \cite{penning2019dtn,sommer2025quicl,iondtn,mud3tn,10.1109/GLOCOM.2018.8647274}. 

Much of what is labeled ``implementation'' in teaching and survey settings is still \emph{not} something a learner can open, run, and probe. That leaves a gap between \emph{knowing} that DTN exists and \emph{understanding} how orbital motion, RF margin, custody transfer, fragmentation, and end-to-end security interact when a bundle actually moves through the network. The tool we present closes that gap by releasing the entire pipeline, as open source. Every layer is intentionally readable and editable so that users can correlate textbook concepts with concrete functions, experiment with parameters, and extend the system; this transparency is difficult to achieve with theory-only materials or with monolithic, closed implementations.

This motivated the development of our open-source ISS communication simulator \cite{ISSsim-repo}.

\subsection{Learning Outcomes}

This project addresses several pedagogical objectives in the domain of computer networks and communications:

\begin{enumerate}
    \item \textbf{Protocol Understanding:} Students can observe how the Bundle Protocol handles message encapsulation, fragmentation, and custody transfer, concepts that are difficult to appreciate through static diagrams alone.
    
    \item \textbf{Network Simulation:} By leveraging Mininet~\cite{lantz2010network}, students experience realistic network emulation with configurable bandwidth, latency, and packet loss parameters that dynamically reflect orbital geometry.
    
    \item \textbf{Physical Layer Awareness:} The link budget calculations expose students to RF engineering concepts including free-space path loss, atmospheric attenuation, Doppler shift, and Shannon-Hartley capacity limits.
    
    \item \textbf{Security Mechanisms:} The implementation of Bundle Security Protocol (BSP) blocks demonstrates end-to-end security in store-and-forward networks, where intermediate nodes must be trusted custodians.
    
    \item \textbf{Full-Stack Development:} The combination of a Python/FastAPI backend with a React/Three.js frontend showcases modern software engineering practices for building interactive network visualization tools.
\end{enumerate}

\subsection{Relation to Computer Networking}

This work intersects with several core computer networking topics:

\paragraph{Delay-Tolerant Networking.} The DTN architecture~\cite{fall2003delay,cerf2007rfc,warthman2012delay} represents a significant departure from traditional Internet assumptions. Our implementation demonstrates key DTN concepts including the Bundle Protocol's store-and-forward paradigm, custody transfer mechanisms for reliable delivery, and routing for path computation in networks with scheduled connectivity. The simulator visualizes how bundles traverse a mesh of ground stations, waiting at intermediate nodes until ISS contact windows enable final delivery.

\paragraph{Network Emulation with Mininet.} Mininet~\cite{lantz2010network} enables the creation of realistic virtual network topologies with software-defined networking capabilities. Our system creates a partial mesh of ground station nodes connected to a central ISS node, with link parameters (bandwidth, delay, packet loss) dynamically updated based on real-time orbital calculations. This allows students to observe actual packet flows rather than simulated abstractions.

\paragraph{Link Budget and Physical Layer.} While computer networking courses typically focus on layers 3 and above, space communications require understanding the physical layer constraints that determine link availability. Our link budget calculator implements the complete RF path analysis: transmit power, antenna gains, free-space path loss (FSPL), atmospheric attenuation as a function of elevation angle, and thermal noise calculations. The resulting signal-to-noise ratio (SNR) determines achievable data rates through Shannon-Hartley capacity estimation, directly affecting bundle transmission times visualized in the interface.

\paragraph{Routing in Challenged Networks.} Unlike conventional routing, DTN routing must account for time-varying connectivity. Our implementation uses live visibility and next-pass predictions for routing decisions. Bundles are forwarded through the ground station mesh toward stations with earlier predicted contact windows, demonstrating opportunistic routing strategies absent from traditional networking curricula.

This paper is organized as follows:
in Sec.~\ref{sec:methods}, we describe the implementation details and tools
employed in building our framework, discussing the different architectural components
and calculations;
Sec.~\ref{sec:usecase}, presents the educational use cases and practical applications
of the platform, integrated learning tools and
how to best use the platform;
Sec.~\ref{sec:eval} provides a quantitative evaluation of the simulator, reporting
experimental results on bundle delivery performance, security overhead,
fragmentation behavior, scalability, and Mininet-based network emulation validation;
Sec.~\ref{sec:concl} finalizes by drawing some concluding remarks
and plausible areas of future work.

%%%%%%%%%%%%%%%%%%%%%%%%%%%%%%%%%%%%%%%%%%%%%%%%%%%%%%%%%%%
\section{Methodology: Implementation Details}
\label{sec:methods}
%%%%%%%%%%%%%%%%%%%%%%%%%%%%%%%%%%%%%%%%%%%%%%%%%%%%%%%%%%%

The open-source repository is 
accessible at
	\url{https://github.com/kritgrover/iss-simulator}.
Further documentation, technical details, tutorials and video-demonstrations
are also available in the repository.
A testable deployment has been made available at,
	\url{https://utsc.utoronto.ca/webapps/iss-simulator/}
that offers a way for users to test the basic functionalities of
the system.\footnote{Because users will be mostly interacting with
the frontend some of the underlying backend features are not accessible.}

\subsection{DTN Architecture and Bundle Protocol Overview}

The Delay-Tolerant Networking architecture fundamentally differs from the traditional OSI model by introducing a new protocol layer, the Bundle Protocol, that operates as an overlay above the transport layer. In our implementation, bundles are self-contained messages that include source and destination endpoints, payload data, security blocks, and metadata for routing and custody transfer.

The Bundle Protocol (BP) as specified in RFC~5050 \cite{scott2007rfc} and RFC~9171 \cite{10.17487/RFC9171} defines bundles as the fundamental unit of data transmission in DTN networks. Each bundle contains a primary block (routing and metadata), payload block (application data), and extension blocks for security, fragmentation, and other services. Our implementation adheres to this structure while providing both simulation and network emulation modes to demonstrate protocol behavior under realistic conditions.

The system is built using a modern tech stack integrating the following
elements:
\begin{itemize}
    \item{\textbf{Backend}:
		Python with FastAPI \cite{lubanovic2023fastapi} and WebSocket \cite{pimentel2012communicating} server,
		utilizing Skyfield \cite{2019ascl.soft07024R} for orbital tracking
		using Simplified General Perturbations model (SGP4) \cite{doi:10.2514/6.2008-6770},
		and Mininet (optional) \cite{lantz2010network} for realistic network emulation.}
		
	\item{\textbf{Frontend}:
		React (Vite) with TypeScript \cite{sakhniuk2024react}, using Three.js \cite{Danchilla2012} for 3D globe visualization,
		Recharts \cite{elrom2021integrating} for real-time analytics, and Tailwind CSS \cite{gerchev2022tailwind} for a responsive UI.}

	\item{\textbf{Protocols}:
		Implements a simulation of the Bundle Protocol (RFC~5050/9171) \cite{scott2007rfc,10.17487/RFC9171}
		with support for custody transfer, fragmentation, priority queuing
		and encryption.}
\end{itemize}

\subsection{System Architecture Overview}

The simulator employs a client-server architecture with clear separation between backend simulation logic and frontend visualization. The backend operates in two distinct modes: \textit{simulation mode} and \textit{Mininet emulation mode}, selectable via the \texttt{USE\_MININET} environment variable. In simulation mode, bundle transmission is modeled through time-based calculations without actual network sockets, enabling rapid prototyping and educational demonstrations. Mininet mode creates a virtual network topology with real TCP/IP sockets, allowing packet-level inspection with tools like Wireshark \cite{ndatinya2015network} and providing authentic network behavior including actual packet loss, latency, and bandwidth constraints. The backend consists of several core modules: 
\begin{itemize}
    \item \texttt{orbital\_tracker.py} handles SGP4-based position calculations
    \item \texttt{link\_budget\_calculator.py} computes RF link parameters
    \item \texttt{dtn\_bundle\_manager.py} implements the Bundle Protocol logic
    \item \texttt{bsp\_security.py} provides BSP encryption and authentication
    \item \texttt{bundle\_fragmentation.py} manages Maximum Transmission Unit (MTU)-based fragmentation
    \item \texttt{network\_dtn\_manager.py} extends the bundle manager for Mininet network operations
\end{itemize}
A SQLite database persists bundle metadata, transmission history, and transfer records for analysis and debugging.

The frontend communicates with the backend through WebSocket connections for real-time updates and REST API endpoints for bundle creation and management. State management follows React's component-based architecture with custom hooks encapsulating data fetching and WebSocket communication logic.

%%%%%%%%%%%%%%%%%%%%%%%%%%%%%%%%%%%%%%%%%%%%%%%%%%%%%%%%%%%%
\subsection{Frontend Implementation}
\label{sec:frontend}
%%%%%%%%%%%%%%%%%%%%%%%%%%%%%%%%%%%%%%%%%%%%%%%%%%%%%%%%%%%%

The frontend is built as a single-page application using React with TypeScript, providing real-time visualization of orbital mechanics, bundle transmission, and network topology. The interface is organized into two main views: the \textit{Ground View} (main interface) and the \textit{ISS View} (dedicated ISS perspective for relaying back and decrypting messages).

\subsubsection{3D Globe Visualization}

\begin{figure*}
	\includegraphics[width=\textwidth]{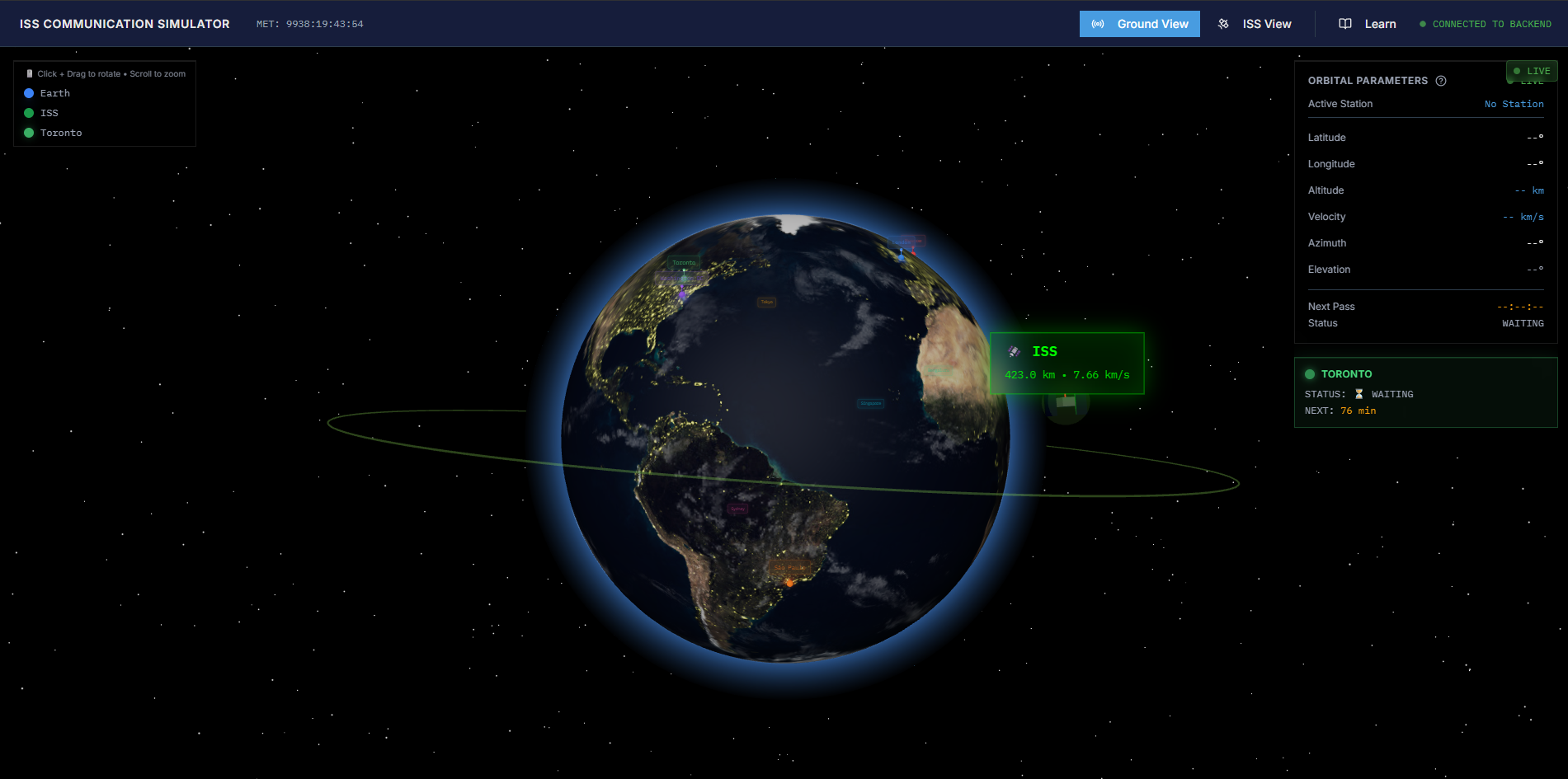}
	\caption{Main dashboard interface showing 3D globe visualization, orbital parameters, and next pass.}
	\label{fig:dashboard-globe}
\end{figure*}

The 3D globe component uses Three.js to render an interactive Earth model with real-time ISS position and orbital path visualization, as shown in Fig.~\ref{fig:dashboard-globe}.
The implementation employs WebGL for hardware-accelerated rendering, ensuring smooth 60 FPS performance even with complex geometries.

The globe is constructed from a high-resolution Earth texture map (NASA Blue Marble) applied to a sphere geometry.  The ISS is represented as a 3D model positioned using a spherical-to-cartesian coordinate transformation with coordinate system adjustments.
These relations are what allow the frontend to \emph{track} the vehicle in a single consistent frame: geodetic latitude, longitude, and altitude from the backend are converted to scene coordinates each update cycle so the ISS sprite, ground-station markers, and ``visible'' link rays stay aligned as orbital state changes. Without an explicit mapping, small Earth models and WebGL axes would drift relative to the texture and station pins; the equations enforce that the same numeric state drives both the analytics panels and the 3D view.

\begin{equation}
\begin{aligned}
	\phi &= (90° - \text{lat}) \cdot \frac{\pi}{180} \quad \text{(colatitude)} \\
	\theta &= (\text{lon} + 180°) \cdot \frac{\pi}{180} \quad \text{(longitude with offset)} \\
	x &= -(R + h) \sin(\phi) \cos(\theta) \\
	z &= (R + h) \sin(\phi) \sin(\theta) \\
	y &= (R + h) \cos(\phi)
\end{aligned}
\end{equation}
where $R$ is Earth radius, $h$ is altitude, $\phi$ is colatitude, and $\theta$ is longitude with a 180° offset. The transformation maps to a Three.js coordinate system where the y-axis points up, and the x-axis is negated to align with the texture mapping.

%%%%%%%%%%%%%%%%%%%%%%%%%%%%%%%%%%%%%%%%%%%%%%%%%%%%%%%%%%%%
\subsection{Backend Engine}
\label{sec:backend}
%%%%%%%%%%%%%%%%%%%%%%%%%%%%%%%%%%%%%%%%%%%%%%%%%%%%%%%%%%%%

\subsubsection{Orbital Tracking and Position Calculation}

The orbital tracking subsystem uses the Simplified General Perturbations 4 (SGP4) propagation model to compute the ISS position in real-time. Two-Line Element (TLE) data is fetched from \texttt{celestrak.org} or from locally cached previous celestrak call, providing the orbital parameters necessary for SGP4 calculations. The \texttt{OrbitalTracker} class, built on the Skyfield library, computes the ISS position in Earth-Centered Inertial coordinates and transforms it to geodetic coordinates for visualization.

For each ground station, the system calculates look angles using spherical trigonometry. The elevation angle determines visibility: a station is considered visible when elevation exceeds a configurable threshold (default 0°). Pass prediction algorithms identify upcoming contact windows by searching for elevation transitions above the threshold, computing acquisition of signal and loss of signal times at 1 minute intervals.

\subsubsection{Link Budget Calculations}

We implemented a comprehensive RF link analysis following standard space communications engineering practices. As the orbital tracker updates ISS position and each ground station's look angles (range, elevation, azimuth), the calculator refreshes the same RF quantities a mission analyst would monitor: whether the link is physically plausible, how much margin exists over noise, and most importantly, the \emph{data rate} that can be sustained during a pass. Those tracked values feed the UI (SNR, capacity, estimated transfer time) and the transmission scheduler: a higher effective rate shortens in-contact bundle delivery, while a loss of margin corresponds to a loss of link in the dynamics. 

The link budget computes signal-to-noise ratio through the following chain:

\begin{equation}
\text{SNR}_{\text{dB}} = P_t + G_t + G_r - L_{\text{FS}} - L_{\text{atm}} - L_{\text{cable}} - L_{\text{misc}} - N_{\text{floor}}
\end{equation}

where $P_t$ is transmit power, $G_t$ and $G_r$ are transmit and receive antenna gains, $L_{\text{FS}}$ is free-space path loss, $L_{\text{atm}}$ is atmospheric attenuation, $L_{\text{cable}}$ and $L_{\text{misc}}$ are cable and miscellaneous system losses, and $N_{\text{floor}}$ is the noise floor power. Together, these terms explain \emph{why} SNR changes as range and elevation change when tracking the pass. The noise floor is calculated as:

\begin{equation}
N_{\text{floor}} = 10\log_{10}(k_B T B \times 1000)
\end{equation}

where $k_B$ is Boltzmann's constant, $T$ is system noise temperature, and $B$ is bandwidth. Tracking $N_{\text{floor}}$ fixes the sensitivity baseline against which received power is compared as geometry evolves.

Free-space path loss is calculated using the Friis transmission equation:
\begin{equation}
L_{\text{FS}} = 20\log_{10}\left(\frac{4\pi d f}{c}\right)
\end{equation}
where $d$ is range in kilometers, $f$ is frequency in MHz, and $c$ is the speed of light. Since $d$ comes directly from the instantaneous ISS station geometry, $L_{\text{FS}}$ is the dominant lever when the user follows a pass from rise to set.

Atmospheric attenuation varies with elevation angle due to increased path length through the troposphere at low angles. Our implementation uses an empirical model with a path length scaling factor:
\begin{equation}
L_{\text{atm}} = L_0 \cdot \min\left(\frac{1}{\sin(\theta)}, 10.0\right)
\end{equation}
where $\theta$ is elevation angle in radians and $L_0 = 0.5$ dB is the attenuation at zenith. The path length factor is capped at 10.0 to prevent unrealistic losses at very low elevation angles. Thus, as the dashboard tracks elevation through a contact, $L_{\text{atm}}$ explains additional fade near the horizon that pure free-space models omit.

Doppler shift compensation is critical for maintaining carrier lock. The relative velocity between ISS and ground station is computed from the orbital state vector, and Doppler shift is calculated as:
\begin{equation}
f_{\text{Doppler}} = f_0 \frac{v_r}{c}
\end{equation}
where $f_0$ is the carrier frequency and $v_r$ is the radial velocity component in km/s. Tracking $f_{\text{Doppler}}$ connects the kinematic state from SGP4 to a physical-layer concern students can relate to frequency stability and receiver design, even when the pedagogical UI stresses rate and margin before carrier-loop details.

The achievable data rate is determined using the Shannon-Hartley theorem:
\begin{equation}
C = B \log_2(1 + \text{SNR}_{\text{linear}})
\end{equation}
where $C$ is channel capacity in bits per second. The system applies a 75\% efficiency factor to account for coding and modulation inefficiencies, yielding practical data rates that directly influence bundle transmission times. In other words, the quantity the user sees as ``Mbps'' or transfer ETA is traced from the same SNR chain above; tracking $C$ over time makes concrete how orbital motion and weather-dominated losses bound DTN throughput during short contact windows.

\subsubsection{Bundle Protocol Implementation}

The \texttt{DTNBundleManager} class implements the core Bundle Protocol functionality. Each bundle is represented as a \texttt{DTNBundle} dataclass containing: bundle identifier, source and destination endpoints, encrypted payload, priority level, creation timestamp, time-to-live, custody transfer flag, hop list for loop prevention, and security blocks.

\begin{figure*}[h]
	\includegraphics[width=\textwidth]{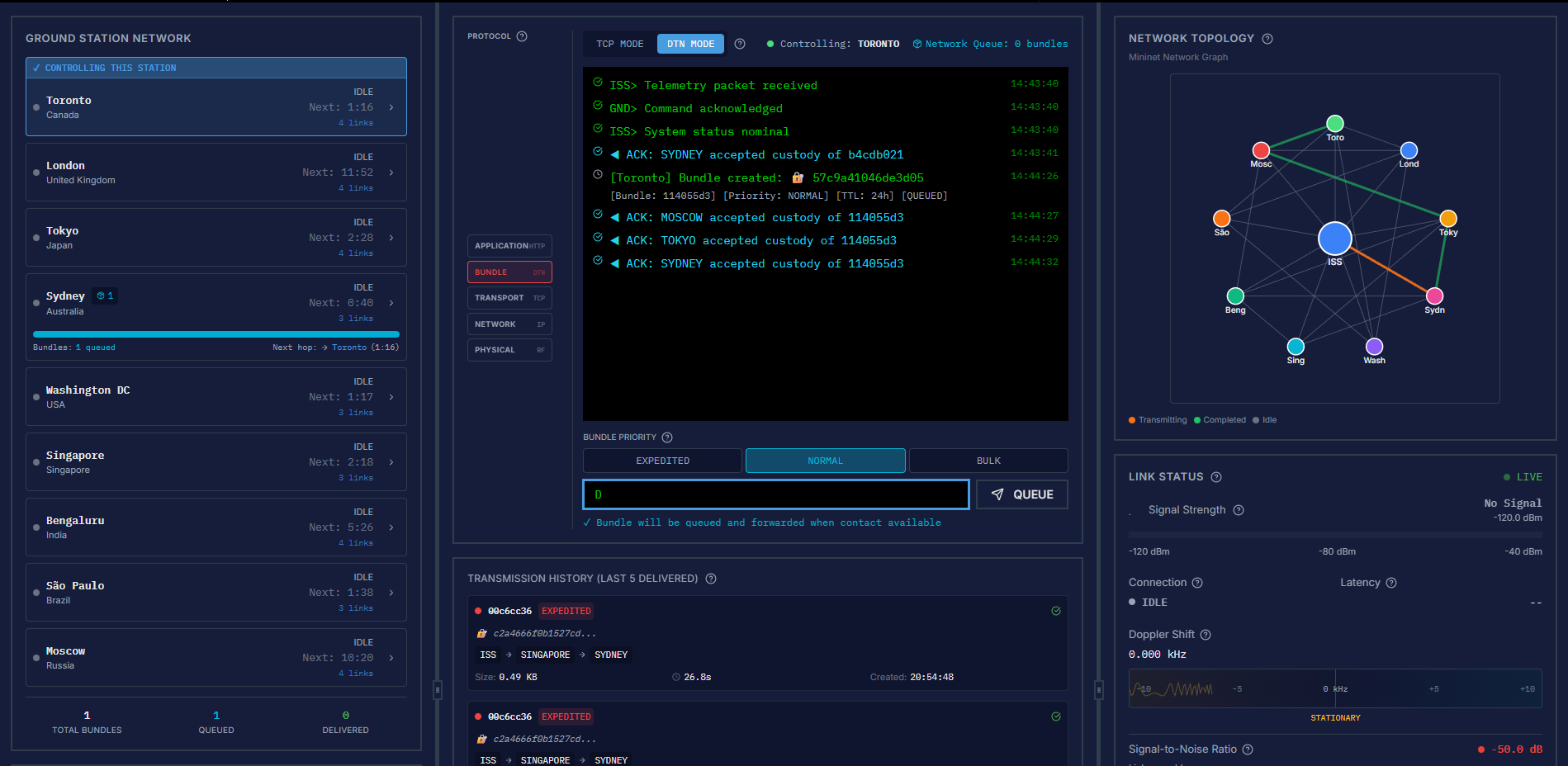}
	\caption{Ground View showing the interface for creating bundles that show the encrypted text, network topology, link parameters and transmission history.}
	\label{fig:message-interface}
\end{figure*}

Bundle creation follows the BP specification: upon receiving a message from a ground station, the system generates a unique bundle ID, encrypts the payload using Bundle Security Protocol, computes a SHA-256 payload hash for integrity verification, and initializes routing metadata. The bundle is then enqueued at the source station's transmission queue, sorted by priority and creation time.

Priority queuing ensures expedited bundles are transmitted first during contact windows. The queue sorting algorithm uses a composite key: priority level (EXPEDITED $>$ NORMAL $>$ BULK), then creation timestamp. This guarantees fair ordering within priority classes while ensuring time-critical messages receive preferential treatment.
Fig.~\ref{fig:message-interface} shows the message interface including several aspects of the bundles logistics.

\subsubsection{Bundle Security Protocol (BSP)}

The Bundle Security Protocol implementation in \texttt{bsp\_security.py} provides three security block types as specified in RFC~6257 \cite{rfc6257}:

\textbf{Bundle Authentication Block (BAB):} Provides hop-by-hop authentication between nodes. Each BAB is created using HMAC-SHA256 with a shared secret key, authenticating the bundle's integrity and source at each forwarding node. The BAB includes the security source, security destination (next hop), and a cryptographic signature over the bundle's metadata and payload hash. BABs are regenerated at each hop to reflect the new security source-destination pair, ensuring that intermediate nodes cannot replay or modify bundles without detection.

\textbf{Payload Integrity Block (PIB):} Provides end-to-end integrity verification of the bundle payload. The PIB contains a signature computed over the encrypted payload hash. The PIB is created after encryption, ensuring that the integrity check covers the ciphertext that will be transmitted. Upon delivery, the destination verifies payload integrity by recomputing the HMAC signature over the encrypted payload hash and comparing it against the PIB signature value. This ensures that the encrypted payload received matches what was originally signed, without requiring decryption for integrity verification.

\textbf{Payload Confidentiality Block (PCB):} Provides end-to-end encryption of the bundle payload using AES-256-CBC. The encryption key is derived from a shared secret using PBKDF2 with 100,000 iterations and SHA-256. The PCB includes the initialization vector and ciphertext, allowing only the destination node (which has the shared key) to decrypt the payload. Intermediate nodes can forward encrypted bundles without access to plaintext, providing confidentiality even in untrusted network environments.

The security block creation process follows a specific order: PIB and PCB are created at bundle creation, then BAB is created at each hop during transmission. This ordering ensures that BAB is the outermost security layer, providing hop-by-hop authentication of the bundle metadata and payload hash, while PIB and PCB provide end-to-end integrity and confidentiality protection.

\subsubsection{Bundle Fragmentation and Reassembly}

Large bundles exceeding the Maximum Transmission Unit (MTU) must be fragmented to traverse heterogeneous network links. We implemented the fragmentation procedures with a configurable MTU of 4096 bytes, including headers.

\begin{figure}
 	\includegraphics[width=\columnwidth]{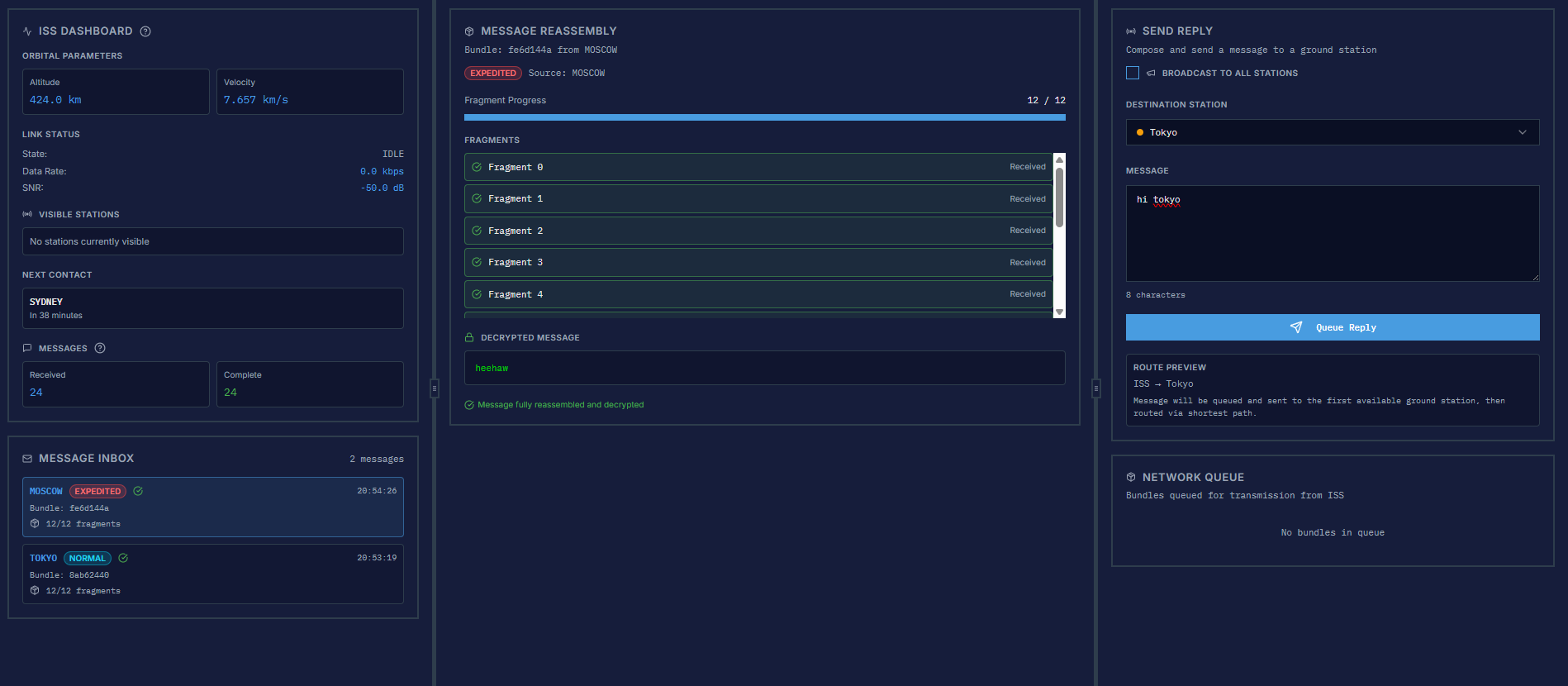}
 	\caption{ISS View showing a relay interface, message inbox, and a reassembly component.}
 	\label{fig:iss-view}
 \end{figure}

Fragmentation occurs automatically when a bundle's total size exceeds the MTU. The fragmentation algorithm divides the encrypted payload into fixed-size chunks, each assigned a fragment number starting from 0. Each fragment becomes an independent bundle with its own bundle ID, but shares the parent bundle ID in metadata for reassembly tracking. An example of this is shown in Fig.~\ref{fig:iss-view}.

Fragment bundles include: parent bundle ID, fragment number, total fragment count, and the fragment payload. All fragments share the same security blocks from the parent bundle, ensuring end-to-end security is preserved across fragmentation boundaries.

Reassembly occurs at the destination node. The \texttt{reassembly\_buffers} dictionary tracks received fragments by parent bundle ID. When all fragments for a parent bundle are received, the fragments are sorted by fragment number and concatenated to reconstruct the original encrypted payload. The reassembled payload is then decrypted. PIB integrity verification occurs when each fragment arrives, ensuring the reassembled encrypted payload will match the original before decryption.

\subsubsection{Routing Algorithms}

Our implementation uses a next-pass prediction approach that considers predicted ISS passes over ground stations, to decide which station to forward the bundle to. We implemented a Breadth-First Search path-finding algorithm that builds routes through the ground station mesh. For bundles originating at ground stations destined for ISS, the algorithm:

\begin{enumerate}
	\item Identifies stations with ISS visibility, and selects the station with the earliest predicted contact window
	\item If the selected station is not the source, uses BFS to find a path to that station through the mesh network
	\item Returns the complete route: [source, intermediate\_stations..., contact\_station, ISS]
\end{enumerate}

For bundles originating at ISS destined for ground stations, the algorithm reverses: it finds currently visible stations or stations with soonest upcoming passes, then routes through the mesh to the final destination.

The mesh topology is defined by pairwise connections between ground stations, creating a partial mesh network. Ground station links operate at 100 Mbps with minimal latency (simulated), enabling rapid bundle forwarding between stations. The routing algorithm avoids loops by maintaining a visited node list and checking the bundle's hop history before forwarding.

Broadcast bundles use a flooding algorithm: the bundle is forwarded to all connected stations, with duplicate detection preventing infinite loops. Each station maintains a \texttt{broadcast\_received} set tracking which bundles have already been processed.

\subsubsection{Custody Transfer and Reliability}

\begin{figure}
 	\includegraphics[width=\columnwidth]{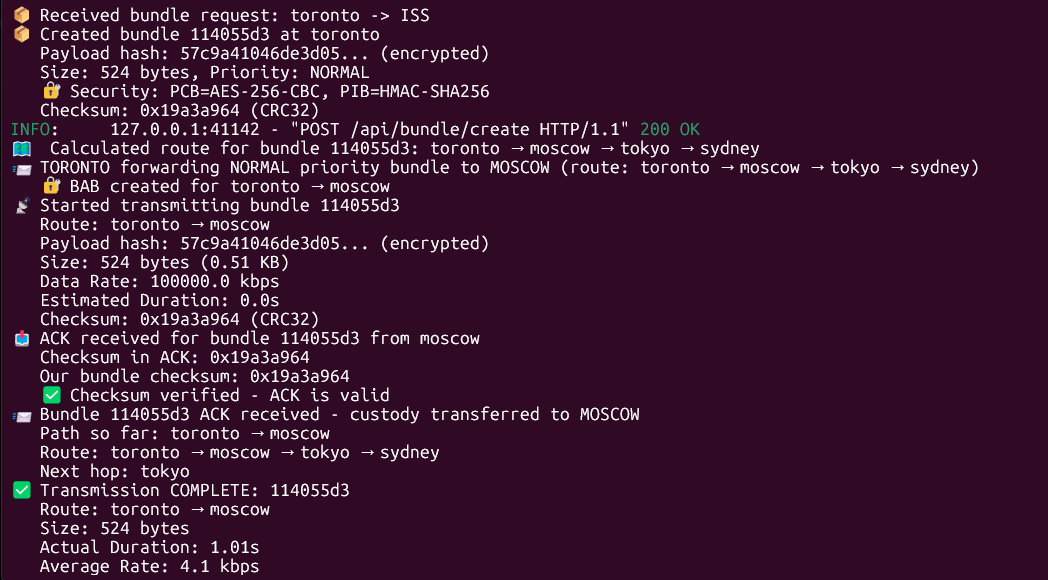}
	 \caption{Backend logs showing routing and received ACKs, checksum and other information for tracing the bundle (Mininet Mode)}
 	\label{fig:backend-terminal}
 \end{figure}

Custody transfer provides reliable delivery guarantees in DTN networks. When a bundle is transmitted with custody transfer enabled, the receiving node becomes responsible for delivery, sending a custody acknowledgment (ACK) or negative acknowledgment (NAK) back to the sender. Fig.~\ref{fig:backend-terminal} shows a scenario like this.

Our implementation tracks pending acknowledgments with timeout mechanisms. The \texttt{PendingAcknowledgment} class stores: bundle ID, expected ACK source, timeout timestamp, and retransmission count. If an ACK is not received within 30 seconds, the system marks the transmission as failed and retries up to 5 times, after which the bundle is dropped.

Delivery confirmation occurs when a bundle reaches its final destination. The destination node sends a delivery ACK, and the bundle status is updated to \texttt{DELIVERED}. Delivered bundles are logged to the database for historical analysis and removed from active transmission queues.

\subsubsection{Thread Safety and Concurrency}

The backend operates in a multi-threaded environment: the main FastAPI event loop handles WebSocket connections and HTTP requests, while background threads manage bundle transmission, Mininet network operations, and database writes. Thread safety is critical to prevent race conditions when multiple threads access shared bundle state.

The \texttt{DTNBundleManager} relies on Python's Global Interpreter Lock (GIL) for basic thread safety. However, the GIL only ensures atomicity of individual bytecode operations and does not prevent race conditions when multiple threads modify shared data structures like dictionaries and lists. Bundle dictionaries, station queues, and the ISS queue are accessed without explicit locking, which means concurrent modifications could cause inconsistencies. Most operations occur within the FastAPI event loop or are serialized through the event loop's single-threaded nature, but explicit locking would be required for true thread safety.

The \texttt{NetworkDTNManager} (Mininet mode) implements per-node locks using \texttt{threading.Lock} objects stored in a \texttt{\_node\_locks} dictionary. These locks are specifically used to serialize \texttt{node.cmd()} calls, as Mininet's command execution is not thread-safe and concurrent calls can cause \texttt{AssertionError} exceptions:

\begin{lstlisting}[frame=single, basicstyle=\footnotesize, language=python]
node_lock = self._get_node_lock(from_node)
with node_lock:
    # Only node.cmd() calls are protected
    result = source_node.cmd(cmd)
\end{lstlisting}

This fine-grained locking allows concurrent operations on different nodes while preventing conflicts when multiple threads execute commands on the same node.

Database operations use SQLite's built-in thread safety mechanisms. The \texttt{DatabaseManager} creates a connection per method call (not per thread), and each connection is closed immediately after use. Write operations are serialized through SQLite's internal locking, which prevents database corruption while allowing concurrent reads. This approach ensures thread safety at the database level without requiring explicit application-level locking.

\subsubsection{Simulation Mode vs. Mininet Emulation Mode}

The system supports two operational modes with distinct characteristics:

\textbf{Simulation Mode:} Bundle transmission is modeled through time-based calculations without actual network sockets. When a bundle transmission is initiated, the system calculates transmission duration based on bundle size and data rate, then tracks progress using periodic time-based updates in the main loop. This mode provides fast execution, deterministic behavior, and easy debugging, making it ideal for educational demonstrations and rapid prototyping. However, it does not capture real network effects like packet loss, actual TCP/IP behavior, or network congestion.

\textbf{Mininet Emulation Mode:} Creates a virtual network topology using Mininet's software-defined networking capabilities. The topology consists of: one ISS node (IP: 10.0.0.100/24), multiple ground station nodes (IPs: 10.0.0.1-9/24), and an OpenVSwitch connecting ground stations in a partial mesh. Bundle transmission occurs over real TCP sockets between Mininet nodes, with link parameters (bandwidth, delay, packet loss) dynamically updated based on orbital calculations.

In Mininet mode, the \texttt{NetworkDTNManager} extends \texttt{DTNBundleManager} to handle socket-based communication. Each ground station runs a DTN server listening on port 5000, accepting bundle transmissions from other nodes. The \texttt{LinkParameterManager} updates Mininet link parameters in real-time: when ISS becomes visible to a station, the link bandwidth increases and delay decreases; when visibility is lost, the link is effectively disabled, by setting the link bandwidth to a very low value.

Mininet mode enables packet-level inspection with Wireshark, allowing students to observe actual DTN protocol messages, encryption headers, and network behavior. This provides authentic network emulation but requires root privileges and more computational resources.

The system automatically falls back to simulation mode if Mininet is unavailable or initialization fails, ensuring robust operation across different deployment environments.

\subsubsection{Database Persistence}

Bundle metadata and transmission history are persisted to a SQLite database for analysis, debugging, and recovery. The \texttt{DatabaseManager} class provides thread-safe database operations with automatic schema migration and corruption recovery.

The database schema stores complete bundle state including: bundle identifiers, source and destination endpoints, encrypted payload (base64-encoded), payload hash for integrity verification, priority and status, creation and delivery timestamps, custody transfer information, hop list and routing path, security blocks, and fragmentation metadata.

Complex data structures are serialized to JSON strings for storage, then deserialized upon retrieval. This approach balances schema simplicity with data flexibility, allowing the system to store arbitrary routing paths and security block structures without schema changes.

The database implements automatic migration: when new columns are added, the migration function detects missing columns and adds them without data loss. This ensures backward compatibility with existing database files while supporting new features.

Corruption recovery is handled gracefully: if database corruption is detected, the system automatically recreates the database file and reinitializes the schema. While this results in data loss, it prevents system crashes and allows continued operation. In production deployments, regular backups would mitigate this risk.

Database operations rely on SQLite's internal locking for concurrent access. Each database method performs operations synchronously, i.e, each method completes its transaction before returning. SQLite's internal locking handles concurrent reads and writes, preventing database corruption from simultaneous access. While this provides basic thread safety for SQLite operations, the application does not implement explicit application-level locks. For high-concurrency scenarios, additional synchronization at the application level may be needed to coordinate complex multi-step operations.

%%%%%%%%%%%%%%%%%%%%%%%%%%%%%%%%%%%%%%%%%%%%%%%%%%%%%%%%%%%
\section{Use Cases and Practical Applications}
\label{sec:usecase}
%%%%%%%%%%%%%%%%%%%%%%%%%%%%%%%%%%%%%%%%%%%%%%%%%%%%%%%%%%%

The ISS communication simulator has been developed as an educational platform that bridges theoretical and practical understanding of DTN concepts.
Because of this, it serves as a perfect tool to motivate  
understanding delay-tolerant networking in space communications,
along with practical demonstrations of its use.

The modular codebase structure, comprehensive documentation, and use of standard technologies (Python, React, TypeScript) lower the barrier to entry for contributions, making the project accessible to users with intermediate programming skills. The MIT license encourages sharing of modifications and extensions, fostering a community of learners and contributors.

The simulator has been designed with extensibility in mind, allowing the educational content to grow with the project. Future enhancements could include:

\begin{itemize}
    \item \textbf{Interactive Tutorials:} Step-by-step guided tours that walk users through key concepts while using the simulator, providing structured learning paths for different skill levels.
    
    \item \textbf{Assignment Templates:} Pre-configured scenarios and exercises that instructors can assign, with automated assessment of student solutions (e.g., ``optimize routing to minimize delivery time for a given message set'').
    
    \item \textbf{Comparative Protocol Analysis:} Side-by-side comparison modes showing how traditional TCP/IP would fail in the same scenarios where DTN succeeds, providing visual demonstrations of protocol differences.
    
    \item \textbf{Multi-Satellite Scenarios:} Extensions to support multiple satellites or satellite constellations, enabling studies of more complex space network topologies.
    
    \item \textbf{Performance Analytics:} Enhanced metrics and visualization tools for analyzing routing efficiency, delivery success rates, and network utilization patterns.
\end{itemize}

By making the ISS communication simulator open-source and education-focused, we aim to lower the barrier to understanding delay-tolerant networking and space communications, making these critical technologies accessible to the next generation of network engineers and space systems developers. The combination of theoretical learning resources (tooltips, Learn tab) with practical hands-on exploration (Wireshark, fragmentation observation, encryption workflows) provides a comprehensive educational experience.

%%%%%%%%%%%%%%%%%%%%%%%%%%%%%%%%%%%%%%%%%%%%%%%%%%%%%%%%%%%
\section{Experimental Evaluation}
\label{sec:eval}
%%%%%%%%%%%%%%%%%%%%%%%%%%%%%%%%%%%%%%%%%%%%%%%%%%%%%%%%%%%

To quantitatively characterize the simulator's behavior and validate its correctness, we conducted controlled experiments in the two modes (Simulation and Mininet). Both modes use deterministic contact schedules so runs are reproducible.

\subsection{Experimental Setup}

\paragraph{Simulation mode.}
Experiments E1 and E4--E6 were executed in simulation mode on a standard workstation. The ground station network comprises nine geographically distributed stations (Toronto, London, Tokyo, Sydney, Washington~DC, Singapore, Bengaluru, S\~{a}o~Paulo, Moscow) connected through a mesh topology with 100~Mbps inter-station links.

Contact windows follow a synthetic schedule based on representative ISS orbital parameters: an orbital period of 92~minutes with 8-minute contact windows per station, staggered across stations to simulate the ground-track progression of a 51.6$^\circ$ inclination orbit. During contact, the ISS--ground link operates at 56~kbps (representative of S-band amateur radio links) with a range of approximately 800~km and an elevation angle of 30$^\circ$. The script \texttt{experiment\_runner.py} injects bundles at configured times and advances a virtual clock, compressing long orbital timelines into seconds of wall-clock time. All bundles use AES-256-CBC encryption and HMAC-SHA256 authentication (BAB/PIB/PCB security blocks active).

\paragraph{Mininet emulation mode.}
Experiments E3, E7, and E8 use \texttt{mininet\_experiment\_runner.py}, which instantiates the same nine-station partial-mesh topology as the live system (\texttt{NetworkDTNManager} over Mininet): ground-to-ground links at 100~Mbps and ISS--ground links updated via Linux \texttt{tc} with 56~kbps bandwidth, 3~ms delay, and a configurable random loss percentage. A \emph{wall-clock} schedule toggles all stations between 2-minute contact windows and 3-minute gaps; during ``down'' intervals the ISS links are severely degraded (near-zero effective bandwidth) so that traffic must wait for the next window. Bundle transfers use real TCP connections between Mininet hosts; \texttt{MetricsCollector} records per-hop socket RTT, send success counts, and observed loss in addition to DTN-level delivery statistics.

\paragraph{Reproducibility.}
Complete instructions for installing dependencies, running the backend experiments, and interpreting outputs are maintained in the public repository alongside the code~\cite{ISSsim-repo}.

\subsection{Baseline Delivery and Latency}
\label{sec:eval-baseline}

We first establish baseline performance by transmitting 20 bundles of 500~bytes each from stations distributed across the network to the ISS. Table~\ref{tab:baseline} summarizes the results.

\begin{table}[h]
\centering
\caption{Baseline delivery performance (E1: 20 bundles, 500\,B payloads).}
\label{tab:baseline}
\begin{tabular}{lr}
\hline
\textbf{Metric} & \textbf{Value} \\
\hline
Delivery ratio & 100\% (20/20) \\
Mean latency & 64.7\,s \\
Median latency & 2.0\,s \\
95th-percentile latency & 262.0\,s \\
Max latency & 661.0\,s \\
Mean hop count & 1.95 \\
Retransmissions & 0 \\
Custody ACKs & 19 \\
NAKs & 0 \\
\hline
\end{tabular}
\end{table}

All 20 bundles were successfully delivered to the ISS with zero failures. The latency distribution reveals the characteristic bimodal behavior of DTN: bundles created when their source station (or a nearby station) has ISS visibility are delivered within seconds (median 2.0\,s), while bundles created during contact gaps must wait for the next pass, resulting in latencies up to 661\,s ($\approx$11~minutes). The 95th-percentile latency of 262\,s reflects the inter-pass gap of the synthetic contact schedule. On average, each bundle traverses 1.95~hops (one ground-to-ground custody transfer followed by one ground-to-ISS uplink), consistent with the mesh routing strategy that forwards bundles to the station with the soonest ISS contact.

\subsection{Security Overhead}
\label{sec:eval-security}

A key concern for DTN implementations is the overhead introduced by security mechanisms. We measured the size and processing time overhead of the Bundle Security Protocol across payload sizes ranging from 64\,B to 16\,KB. Table~\ref{tab:security} reports the results.

\begin{table}[h]
\centering
\caption{BSP security overhead by payload size (AES-256-CBC encryption, HMAC-SHA256 authentication).}
\label{tab:security}
\begin{tabular}{rrrrr}
\hline
\textbf{Plaintext} & \textbf{Encrypted} & \textbf{Overhead} & \textbf{Overhead} & \textbf{Total Time} \\
\textbf{(B)} & \textbf{(B)} & \textbf{(B)} & \textbf{(\%)} & \textbf{(ms)} \\
\hline
64 & 108 & 44 & 68.8 & 13.11 \\
128 & 192 & 64 & 50.0 & 0.03 \\
256 & 364 & 108 & 42.2 & 0.03 \\
512 & 704 & 192 & 37.5 & 0.03 \\
1\,024 & 1\,388 & 364 & 35.5 & 0.03 \\
2\,048 & 2\,752 & 704 & 34.4 & 0.03 \\
4\,096 & 5\,484 & 1\,388 & 33.9 & 0.08 \\
8\,192 & 10\,944 & 2\,752 & 33.6 & 0.05 \\
16\,384 & 21\,868 & 5\,484 & 33.5 & 0.07 \\
\hline
\end{tabular}
\end{table}

The size overhead converges to approximately 33.5\% for payloads above 1\,KB, attributable to AES block-cipher padding (PKCS7) and Base64 encoding of ciphertext. For small payloads ($\leq$128\,B), the fixed overhead of the initialization vector and padding dominates, reaching 68.8\% for 64-byte payloads. Processing time remains under 0.1\,ms for all payload sizes after the initial key derivation (PBKDF2 with 100\,000 iterations), demonstrating that BSP security adds negligible computational cost to bundle processing. The first encryption call incurs a one-time key derivation cost of $\approx$13\,ms, which is amortized across subsequent operations.

\subsection{Fragmentation Impact}
\label{sec:eval-frag}

Bundles exceeding the Maximum Transmission Unit (MTU) of 2\,048 bytes are automatically fragmented. We evaluated three payload sizes: 1\,KB (below MTU), 4\,KB (2 fragments), and 16\,KB (8+ fragments), each with 10 bundles.

\begin{table}[h]
\centering
\caption{Fragmentation impact on delivery and overhead.}
\label{tab:fragmentation}
\begin{tabular}{lrrr}
\hline
\textbf{Payload} & \textbf{1\,KB} & \textbf{4\,KB} & \textbf{16\,KB} \\
\hline
Delivery ratio & 100\% & 100\% & 100\% \\
Mean latency (s) & 103.5 & 103.5 & 103.5 \\
Security overhead (B) & 844 & 1\,856 & 5\,856 \\
Avg hops & 1.8 & 1.8 & 1.8 \\
Wall time (s) & 0.7 & 1.9 & 6.2 \\
\hline
\end{tabular}
\end{table}

All payload sizes achieved 100\% delivery with identical latency characteristics, demonstrating that the fragmentation and reassembly mechanisms preserve delivery reliability. The security overhead scales proportionally with payload size, and wall-clock processing time increases for larger payloads due to the additional fragment bundles traversing the network. Notably, the 16\,KB payloads produced bundles requiring 8+ fragments each, yet the system correctly reassembled and delivered all of them.

\subsection{Scalability}
\label{sec:eval-scale}

To evaluate system behavior under increasing load, we varied the number of concurrent bundles from 1 to 50 while keeping payload size constant at 500\,B. Table~\ref{tab:scalability} summarizes the results.

\begin{table}[h]
\centering
\caption{Scalability: performance vs.\ number of concurrent bundles.}
\label{tab:scalability}
\begin{tabular}{lrrrrr}
\hline
\textbf{Bundles} & \textbf{1} & \textbf{5} & \textbf{10} & \textbf{25} & \textbf{50} \\
\hline
Delivery ratio & 100\% & 100\% & 100\% & 100\% & 100\% \\
Mean latency (s) & 661.0 & 145.8 & 103.5 & 43.4 & 35.4 \\
P95 latency (s) & 661.0 & 577.2 & 472.0 & 135.4 & 134.8 \\
Custody ACKs & 1 & 5 & 8 & 21 & 44 \\
Avg hops & 2.0 & 2.0 & 1.8 & 1.84 & 1.88 \\
\hline
\end{tabular}
\end{table}

The system maintains a 100\% delivery ratio across all load levels, demonstrating that the priority queuing and custody transfer mechanisms handle concurrent bundles effectively. Mean latency decreases with more bundles because bundles injected closer to a contact window benefit from immediate transmission; with more bundles spread uniformly across time, a higher fraction coincides with active contact windows. The hop count remains stable at approximately 2 hops regardless of load, indicating that the routing algorithm consistently selects efficient paths. These results confirm that the simulator scales gracefully to at least 50 concurrent bundles without degradation.

\subsection{Mininet Emulation Validation}
\label{sec:eval-mininet}

We complement the simulation sweeps above with Mininet experiments that exercise the same DTN logic over real TCP sockets and kernel networking. Table~\ref{tab:sim-vs-mininet} contrasts the simulation baseline (E1: 92-min orbital period, 8-min staggered windows, virtual time) with the Mininet baseline (E8: global 2-min ``up'' / 3-min ``down'' schedule, real wall-clock time, 10\% baseline ISS-link loss unless noted). Both scenarios use 20 bundles of 500\,B to the ISS. The absolute latencies are not directly comparable across modes because the contact timelines and time bases differ; the key observation is that \emph{both} preserve a 100\% bundle delivery ratio, while Mininet adds observable socket-level behavior.

\begin{table}[h]
\centering
\caption{Simulation baseline (E1) vs.\ Mininet baseline (E8), 20 bundles, 500\,B payloads.}
\label{tab:sim-vs-mininet}
\begin{tabular}{lcc}
\hline
\textbf{Metric} & \textbf{E1 (sim)} & \textbf{E8 (Mininet)} \\
\hline
Delivery ratio & 100\% (20/20) & 100\% (20/20) \\
Mean latency & 64.7\,s & 3.2\,s \\
Median latency & 2.0\,s & 1.9\,s \\
95th-percentile latency & 262.0\,s & 9.1\,s \\
Max latency & 661.0\,s & 16.8\,s \\
Mean hop count & 1.95 & 1.0 \\
Custody ACKs / NAKs & 19 / 0 & 0 / 2 \\
Socket sends (ok / total) & --- & 20 / 22 \\
Mean socket RTT & --- & 2735\,ms \\
Observed net.\ loss (sends) & --- & 9.1\% \\
\hline
\end{tabular}
\end{table}

In E8, two socket-level failures were recovered via DTN retransmission/NAK handling (22 sends for 20 bundles), yielding a network-layer delivery ratio of 90.9\% but 100\% application-layer delivery. Average one-way socket RTT averaged 2.7\,s under the shaped ISS link, reflecting slow bandwidth and queueing.

We swept configured ISS-link loss from 0\% to 30\% (10 bundles each, 90\,s run). Table~\ref{tab:mininet-loss} shows that the \emph{bundle} delivery ratio remained 100\% at every loss level: store-and-forward plus custody semantics absorbed packet loss that would break a naive single-attempt transfer. Socket statistics vary with loss: at 5\% configured loss we observed one NAK and one extra send (90.9\% per-send success) before final delivery; at several higher loss settings all ten sends succeeded in the recorded run (stochastic variation is expected).

\begin{table}[h]
\centering
\caption{Mininet E3: bundle delivery vs.\ configured ISS-link loss (10 bundles each).}
\label{tab:mininet-loss}
\begin{tabular}{rrrrr}
\hline
\textbf{loss (\%)} & \textbf{BDR} & \textbf{Mean lat.\ (s)} & \textbf{Mean RTT (ms)} & \textbf{Sends ok/tot.} \\
\hline
0 & 100\% & 3.0 & 3291 & 10/10 \\
5 & 100\% & 9.4 & 6676 & 10/11 \\
10 & 100\% & 2.1 & 1872 & 10/10 \\
20 & 100\% & 1.4 & 1272 & 10/10 \\
30 & 100\% & 1.1 & 922 & 10/10 \\
\hline
\end{tabular}
\end{table}

Finally, we compared DTN delivery to raw TCP under the same intermittent Mininet topology (E7). One bundle was injected from Toronto to the ISS using the full DTN stack and completed successfully (100\% delivery, $\approx$4.6\,s end-to-end latency in that run). We then issued five independent TCP connections from a host to the ISS listener on port~5000 with the same 500-byte payload while the link schedule toggled; \emph{all five} connections failed to complete transfer (0/5), with errors such as ``No route to host'' as the virtual link moved between up and degraded states. This illustrates a pedagogical point central to the tool: conventional TCP assumes a stable path, whereas DTN is designed to span disconnections by retaining custody until the next contact opportunity.

%%%%%%%%%%%%%%%%%%%%%%%%%%%%%%%%%%%%%%%%%%%%%%%%%%%%%%%%%%%
\section{Conclusions \& Discussion}
\label{sec:concl}
%%%%%%%%%%%%%%%%%%%%%%%%%%%%%%%%%%%%%%%%%%%%%%%%%%%%%%%%%%%

In this paper, we present an open-source implementation of the DTN protocol used
in the ISS communication systems.
The project has been developed to allow practitioners to
extend, modify and adapt its capabilities.
The tool is able to interact and work together with standard tools in the networking industry,
such as, Mininet, Wireshark, etc.
Similarly, the implementation also considers several important elements
in the context of computer networks, such as, fragmentation, encryption, data integrity
and validation, among many others. Experiments show 100\% bundle delivery in both simulation and Mininet under fragmentation, concurrent load, and elevated ISS-link loss, where naive TCP transfers failed on the same intermittent schedule, making it clear that store-and-forward custody is indispensable in these challenged environments.

Our goal is to provide interested parties with a starting point they can build upon and adapt in diverse ways to pursue their own areas of interest.

%%%%%%%%%%%%%%%%%%%%%%%%%%%%%%%%%%%%%%%%%%%%%%%%%%%%%%%%%%%

%\section*{Acknowledgment}
\section*{Acknowledgement}

We are thankful to Tianze Sun (CMS/UTSC) for their help and support in setting up the systems to run the ISS-simulator testable deployment. Use of GenAI: Lovable was used for styling the frontend layout, colors and typography. LLMs were used to structure sentences for clarity and to suggest alternative phrasings for the manuscript. The final manuscript was carefully reviewed and corrected by the authors to ensure precision.

%\section*{References}
\bibliographystyle{IEEEtran}
\bibliography{refs}

% Supplementary Material
\begin{comment}
\appendix
\input{supp_material}
\end{comment}

\end{document}